\documentclass[prl, amsmath,amssymb,aps, reprint, preprintnumbers, nofootinbib,superscriptaddress]
              {revtex4-1}
\usepackage{xcolor}
\usepackage{dsfont}
\usepackage{graphicx}   
\usepackage{dcolumn}    
\usepackage{bm}         
\usepackage{hyperref}
\usepackage{float}
\usepackage{lipsum}
\usepackage{afterpage}
\usepackage{placeins}

\usepackage{float}

\usepackage{array}
\usepackage{comment}
\usepackage{ulem}

\def\beq{\begin{equation}}
\def\eeq{\end{equation}}
\def\barr{\begin{eqnarray}}
\def\earr{\end{eqnarray}}

\usepackage{tikz,xcolor,hyperref}

\definecolor{lime}{HTML}{A6CE39}
\DeclareRobustCommand{\orcidicon}{\hspace{-1mm}
	\begin{tikzpicture}
	\draw[lime, fill=lime] (0,0) 
	circle [radius=0.12] 
	node[white] {{\fontfamily{qag}\selectfont \tiny \,ID}};
	\draw[white, fill=white] (-0.0525,0.095) 
	circle [radius=0.007];
	\end{tikzpicture}
	\hspace{-3mm}
}

\foreach \x in {A, ..., Z}{\expandafter\xdef\csname orcid\x\endcsname{\noexpand\href{https://orcid.org/\csname orcidauthor\x\endcsname}
		{\noexpand\orcidicon}}
}

\begin{document}

\preprint{TIFR/TH/21-21}
\title{Neutrino propagation when mass eigenstates and decay eigenstates mismatch}
\author{Dibya~S.~Chattopadhyay\orcidA{}}
\email{d.s.chattopadhyay@theory.tifr.res.in}
\affiliation{Tata Institute of Fundamental Research, Homi Bhabha Road, Colaba, Mumbai 400005, India}
\author{Kaustav~Chakraborty\orcidB{}}
\email{kaustav@prl.res.in}
\affiliation{Physical Research Laboratory, Navrangpura, Ahmedabad 380009, India}
\author{\\Amol~Dighe\orcidC{}}
\email{amol@theory.tifr.res.in}
\affiliation{Tata Institute of Fundamental Research, Homi Bhabha Road, Colaba, Mumbai 400005, India}
\author{Srubabati~Goswami\orcidD{}}
\email{sruba@prl.res.in}
\affiliation{Physical Research Laboratory, Navrangpura, Ahmedabad 380009, India}
\author{S.~M.~Lakshmi\orcidE{}}
\email{lakshmi.mohan@ncbj.gov.pl}
\affiliation{National Centre for Nuclear Research, Ludwika Pasteura 7, 02-093 Warsaw, Poland}

\date{\today}

\begin{abstract}
  We point out that the Hermitian and anti-Hermitian components of the
  effective Hamiltonian for decaying neutrinos cannot be simultaneously
  diagonalized by unitary transformations for all matter densities.
  We develop a formalism for the two-flavor neutrino propagation through matter of uniform density, for neutrino decay to invisible states.
  Employing a resummation of the Zassenhaus expansion,  
  we obtain compact analytic expressions for neutrino survival and conversion
  probabilities, to first and second order in the ``mismatch parameter'' $\bar{\gamma}$.
\end{abstract}

\maketitle

\noindent {\it Introduction} ---
Neutrino oscillation experiments have unequivocally established that
neutrinos have masses, and their flavors mix. However, data still
allow the possibility of new physics effects at a
sub-leading level. Neutrino decay to lighter invisible states
\cite{Bahcall:1972my,Acker:1991ej,Acker:1993sz} is one such possibility.
Solutions to neutrino anomalies via a combination
of oscillation and decay have been studied \cite{Barger:1998xk,Barger:1999bg,Choubey:1999ir,
  Choubey:2000an,Bandyopadhyay:2001ct, Joshipura:2002fb,Bandyopadhyay:2002qg,
  GonzalezGarcia:2008ru,Berryman:2014qha,Abrahao:2015rba,Picoreti:2015ika,Gomes:2014yua,Choubey:2018cfz}.
Most of these papers have analyzed neutrino oscillation
probabilities in vacuum, taking the mass eigenstates to be identical with
the  decay eigenstates for analytical treatment.
Matter effects, if relevant, have been implemented numerically.

The effective Hamiltonian for neutrino decay is non-Hermitian, with
its Hermitian component corresponding to the energy and the anti-Hermitian
component corresponding to decay. The assumption of identifying the
mass (energy) eigenstates to decay eigenstates is not valid in general.
Indeed, even in vacuum, these two components need not commute, and hence
need not be diagonalizable simultaneously by unitary transformations. 
Even for the special circumstances or models where the mass eigenstates and
decay eigenstates coincide in vacuum, matter effects make
this mismatch inevitable.

The non-Hermitian Hamiltonian itself may be diagonalized by a similarity
transformation employing a non-unitary matrix. Using this principle,
the oscillation probabilities in the two-flavor scenario in vacuum were
approximately calculated in \cite{Berryman:2014yoa}.
A similar exercise has also been performed in
\cite{Ascencio-Sosa:2018lbk}, albeit in the context of visible neutrino
decays in matter, but no compact analytical expressions
for probabilities have been presented. 

In this Letter, we present a novel prescription for computing the neutrino survival or conversion
probabilities for the scenario with simultaneous oscillation
and invisible decay of neutrinos propagating in  matter of uniform density.  
We represent the effective Hamiltonian matrix by ${\mathcal H}_m$, where
\beq
    {\mathcal H}_m = H_m - i \Gamma_m /2 \; .
\eeq
Here $H_m$ and $\Gamma_m$ are Hermitian matrices.
We choose to work in the basis where the Hermitian part of the Hamiltonian
is diagonalized.
This is the same as the basis of neutrino mass eigenstates in matter
in the absence of decay. 
In this basis, $H_m$ is a diagonal matrix whose elements depend on
neutrino mass squared differences, neutrino energy, and Earth matter potential.
The flavor evolution of neutrinos takes the form
\begin{equation}{\label{eq:Generalevolution}}
\nu(t) = e^{-i \mathcal{H}_m t} \nu(0) \; .
\end{equation}

Note that since $[H_m, \Gamma_m] \neq 0$ in general, $\mathcal{H}_m$ is not
a normal matrix, and $e^{- i {\mathcal H}_m t} \neq e^{- i H_m t} e^{- \Gamma_m t/2}$.
Thus, one has to express $e^{-i \mathcal{H}_m t} $ in terms of a chain of
commutators using the inverse Baker-Campbell-Hausdorff (BCH) formula,
also known as the Zassenhaus formula
\cite{Zassenhaus-Wilhelm,Zassenhaus-Casas}.
The standard form of this formula cannot be truncated to a finite number of terms
in the current scenario, therefore we employ a resummation technique using
its series expansion \cite{Kimura:2017xxz}.
The procedure facilitates a perturbative expansion of the neutrino survival
and conversion probabilities,
in terms of a small parameter $\bar{\gamma}$ that characterizes the
mismatch between the eigenstates of $H_m$ and $\Gamma_m$.

Our prescription leads to explicit analytical forms for two-flavor neutrino
probabilities in matter. The probabilities in
vacuum, as well as those calculated by using the assumption of coincident
mass and decay eigenstates, emerge as special cases.
This formulation is completely new, and provides a clear framework for analyzing neutrino decay in vacuum and matter on the same footing.
Moreover, the techniques can be applied to any situation where
quantum mechanical evolution in terms of non-Hermitian Hamiltonian is
to be calculated.

\noindent{\it Formalism} ---
The effective Hamiltonian may be written in the basis of
neutrino mass eigenstates in matter as
\begin{equation}
  \mathcal{H}_m = \left(
  \begin{array}{cc}
    a_1-i b_1 & -\frac{1}{2} i \gamma  e^{i \chi } \\
    -\frac{1}{2} i \gamma  e^{-i \chi } & a_2-i b_2 \\
  \end{array}
  \right),
\label{eq:Hm-def}
\end{equation}
where $a_i, b_i, \gamma, \chi$ are real.
Since $\Gamma_m$ needs to be positive semidefinite,
$b_i \geq 0$ and $\gamma^2 \leq 4 b_1 b_2$.
The sign of $\gamma$ is taken to be positive; this
defines the value of $\chi$ uniquely.
The Hermitian part of this Hamiltonian is diagonal, which is ensured by
the choice of basis.
The anti-Hermitian part is composed of the diagonal
components involving $b_i$, and the off-diagonal components involving $\gamma$.
Note that $b_i = [\Gamma_m]_{ii}/2$, and
$\gamma e^{i\chi} = [\Gamma_m]_{12}= [\Gamma_m]_{21}^*$.

For future convenience, we define the complex parameter
$d_i \equiv a_i - i b_i$, 
the differences
$\Delta_a \equiv  a_2-a_1, ~\Delta_b \equiv b_2 - b_1, ~
\Delta_d \equiv  d_2 - d_1$,
and the dimensionless ratios
\beq
\bar{\gamma} \equiv \frac{\gamma}{|\Delta_d|} \;, \quad
\bar{\Delta}_a \equiv \frac{\Delta_a}{|\Delta_d|} \;, \quad
\bar{\Delta}_b \equiv \frac{\Delta_b}{|\Delta_d|} \; .
\eeq
Then, in terms of the identity matrix ${\mathbb I}$ and
\begin{equation}
  \mathbb{X}\equiv - \frac{i \, \Delta_d \, t}{2} \left(
  \begin{array}{cc}
    -1 & 0\\
    0 & 1
  \end{array}
  \right) \; , \mathbb{Y}\equiv  - \frac{\gamma t}{2}\left(
  \begin{array}{cc}
    0 &  e^{i \chi}\\
    e^{-i \chi} & 0
  \end{array}
  \right) \; ,
  \label{eq:x-y-def}
\end{equation}
one may write
\beq
- i \mathcal{H}_m t =  -\frac{i t}{2}\, (d_1 + d_2) \,\mathbb{I}
+ \mathbb{X} + \mathbb{Y} \; .
\eeq
The commutator of $\mathbb{X}$ and $\mathbb{Y}$ is 
\beq
   {\mathcal L}_\mathbb{X} \mathbb{Y} \equiv
    [\mathbb{X}, \mathbb{Y}]
    = i \frac{\gamma \, \Delta_d \,  t^2}{2} \left(
    \begin{array}{cc}
      0 &  - e^{i \chi}\\
      e^{-i \chi} & 0
    \end{array}
    \right) \; ,
\eeq
which will play a key role in our analysis.

\smallskip

\noindent{\it Zassenhaus expansion} ---
In order to calculate the evolution matrix $e^{-i {\mathcal H}_m t}$,
keeping aside the term proportional to the identity matrix, 
we need to calculate the quantity $e^{\mathbb{X}+\mathbb{Y}}$. This may be written in terms of the Zassenhaus expansion
\cite{Zassenhaus-Wilhelm,Zassenhaus-Casas} as
\begin{equation}{\label{eq:BCH}}
  e^{\mathbb{X}+\mathbb{Y}}= e^\mathbb{X} \, e^\mathbb{Y} \,
  e^{-\frac{1}{2}[\mathbb{X},\mathbb{Y}]} \,
  e^{\frac{1}{6}(2[\mathbb{Y},[\mathbb{X},\mathbb{Y}]]+[\mathbb{X},[\mathbb{X},\mathbb{Y}]])}
  ...
\end{equation}
Note that $|\mathbb{Y}| \sim  \bar{\gamma} |\mathbb{X}|$ and 
${\mathcal L}_\mathbb{X} \mathbb{Y}  \sim \bar{\gamma} |\mathbb{X}|^2$, 
where the absolute sign $(| \cdot |)$ represents a typical nonzero element
in the corresponding matrix. This implies that, in general, for
higher-order commutators,
${\mathcal L}^{k-1}_\mathbb{X} \mathbb{Y} \sim \bar{\gamma} |\mathbb{X}|^k$.
Therefore, it is not possible to truncate the
expansion in eq.~(\ref{eq:BCH}) at any fixed order of commutators.
One needs to collect $O(\gamma^k)$ terms from commutators of all orders by
performing a resummation procedure.
We therefore employ the expression for the Zassenhaus expansion in terms of
a series \cite{Kimura:2017xxz}:
\begin{widetext}
\begin{equation}{\label{eq:Kimura}}
    e^{\mathbb{X}+\mathbb{Y}}=
    \Big(1+\Big.
    \Big.\sum\limits_{p=1}^\infty \sum\limits_{n_1,...,n_p=1}^\infty \dfrac{n_p...n_1}{n_p (n_p+n_{p-1})...(n_p+...+n_1)} \mathcal{Y}_{n_p}... \mathcal{Y}_{n_1} \Big)e^\mathbb{X} \; , 
\end{equation}
\end{widetext}
where $ \mathcal{Y}_{n} =\frac{1}{n!}\mathcal{L}^{n-1}_\mathbb{X} \mathbb{Y}$.

To obtain the expansion up to $O(\bar\gamma)$ and $O(\bar\gamma^2)$,
we need to perform the summation for $p=1$ and $p=1,2$, respectively,
since every ${\mathbb Y}$ comes with a factor of $\bar\gamma$.
Thus for an accuracy of $O(\bar\gamma^2)$, we can truncate
\begin{equation}
  {\label{eq:gammasq}}
  e^{\mathbb{X}+\mathbb{Y}} 
  \approx \Big(1+ \sum\limits_{n_1=1}^\infty \mathcal{Y}_{n_1}
  +\sum\limits_{n_1=1}^\infty \sum\limits_{n_2=1}^\infty \dfrac{n_1}{(n_1+n_2)}
  \mathcal{Y}_{n_2}\mathcal{Y}_{n_1}\Big)e^\mathbb{X} ,
\end{equation}
with the double summation term not needed for $O(\bar\gamma)$
accuracy.
One may use
\begin{equation}
  \mathcal{Y}_n =\frac{1}{n!}(i\Delta_d t)^{n-1} \sigma_3^{n-1} \mathbb{Y}\; 
\label{eq:Yn}
\end{equation}
in order to get closed functional forms for the infinite sums.
Here $\sigma_3$ is the Pauli matrix.

\smallskip

\noindent{\it Neutrino flavor conversions up to
  $O(\bar\gamma)$} ---
The truncation of the right hand side of eq.~(\ref{eq:gammasq}) to
the first summation gives
\begin{equation}
  {\label{eq:massgammaamplitude}}
  e^{\mathbb{X}+\mathbb{Y}}=\left(1+
  \dfrac{\sin(\Delta_{d} t)}{\Delta_{d} t}\mathbb{Y}
  -\dfrac{\cos(\Delta_{d} t)-1}{\Delta_{d} t}i\sigma_3 \mathbb{Y}\right)
  e^\mathbb{X}.
\end{equation}
The amplitude matrix in the mass basis in matter is then
\barr
{\label{eq:massgammaamplitude1}}
\mathcal{A}_m \equiv e^{-i \mathcal{H}_m t}
& = & \left(
\begin{array}{cc}
 e^{-i d_1 t} & -i\frac{\gamma  e^{i \chi } g_-(t)}{\Delta_{d}} \\
-i \frac{\gamma  e^{-i\chi } g_-(t)}{\Delta_{d}} & e^{-i d_2 t} \\
\end{array}
\right) ,
\earr
where the functions $g_\pm(t)$ are defined as
\begin{equation}
  g_\pm(t)=\frac{1}{2}(e^{-i d_2 t}\pm e^{-i d_1 t}) \; .
\end{equation}

The neutrino flavor conversion probability $P_{\beta \alpha}$ for
$\nu_\beta \to \nu_\alpha$ conversion may be obtained by calculating
the flavor conversion amplitude
\begin{equation}
  [\mathcal{A}_f]_{\alpha\beta}
  = \left[ U_m \;e^{-i \mathcal{H}_m t}\;U_m^\dagger \right]_
  {\alpha \beta} \; ,
\end{equation}
and further, $P_{\beta \alpha} = |{\cal A}_{\alpha \beta}|^2$.
In the 2-flavour system, 
\begin{equation}
    U_m =\left(
\begin{array}{cc}
 \cos \theta_m  & \sin \theta_m  \\
 -\sin \theta_m & \cos \theta_m \\
\end{array}
\right)
\end{equation} 
is the unitary rotation matrix.
One can write
\begin{equation}{\label{eq:AmpMatrixFlavourBasisUptogamma}}
    \mathcal{A}_f=\left(
\begin{array}{cc}
 g_-(t) A(\chi) + g_+(t) & g_-(t)B(\chi) \\
 g_-(t)B( -\chi) & -g_-(t) A(\chi) + g_+(t) \\
\end{array}
\right) \; ,
\end{equation}
where $A(\chi)$ and $B(\chi)$ are given in Table~\ref{tab:AmpTableGamma}.
The $\chi$-dependence of $A$ and $B$ is implicit wherever not stated
explicitly.

\begin{table}[]
  \caption{\label{tab:AmpTableGamma}%
    The terms in the amplitude matrix ${\mathcal A}_f$ in the flavor basis
    used in eq.~(\ref{eq:AmpMatrixFlavourBasisUptogamma}),
    calculated up to $O(\bar\gamma)$.}
  \centering
  \begin{tabular}{|c|c|}
    \hline
    Term & Expression \\
    \hline
    \hline
    $A(\chi)\equiv A^{(0)}+\gamma A^{(1)}$& $\begin{aligned}
      \smash[b]{\vphantom{\bigg|}}
      -\cos2\theta_m -i\dfrac{\gamma }{\Delta_d} \sin 2\theta_m \cos\chi
      \smash[t]{\vphantom{\bigg|}}
    \end{aligned}$\\
    \hline
    $B(\chi)\equiv B^{(0)}+\gamma B^{(1)}$& $\begin{aligned}
      \smash[b]{\vphantom{\bigg|}}
      \sin 2\theta_m -i \dfrac{\gamma}{\Delta_d}
      \Big( \cos2\theta_m \cos\chi + i \sin \chi \Big) 
      \smash[t]{\vphantom{\bigg|}}
    \end{aligned}$
    \\
    \hline
  \end{tabular}
\end{table}

The survival probability of a neutrino of flavor $\alpha$ is
\begin{equation}
  {\label{eq:PeeUptogamma}}
  \begin{split}
    P_{\alpha\alpha}=&\frac{e^{-(b_1+b_2)t}}{2}\Big[
    (1 + |A|^2) \cosh(\Delta_b t) \Big.\\
    & +(1 - |A|^2) \cos(\Delta_a t) 
     - 2 \text{Re} (A) \sinh(\Delta_b t) \\
    &\Big. + 2 \text{Im} (A) \sin(\Delta_a t) \Big] \; .
    \end{split}
\end{equation}
The survival probability $P_{\beta\beta}$ for the other flavor may be obtained
from $P_{\alpha\alpha}$ with the replacement $A \to -A$.
The probability for $\nu_\beta \to \nu_\alpha$ conversion is
\begin{equation}{\label{eq:PemuUptogamma}}
  P_{\beta \alpha}=\frac{e^{-(b_1+b_2)t}}{2} |B(\chi)|^2 \, 
  [\cosh(\Delta_b t) -\cos(\Delta_a t)] \; .
\end{equation}
The conversion probability $P_{\alpha \beta}$ is obtained by the replacement
$\chi \to -\chi$.
The explicit expressions for the terms in eqs.~(\ref{eq:PeeUptogamma})
and (\ref{eq:PemuUptogamma})
are given in Table~\ref{tab:ProbTableGamma}.

\begin{table}[t]
\caption{\label{tab:ProbTableGamma}%
  The terms to be used in the probabilities shown in eqn.(\ref{eq:PeeUptogamma})
  and (\ref{eq:PemuUptogamma}), calculated up to $O(\bar\gamma)$.}
\centering
\begin{tabular}{|c|c|}
\hline
Term & Expression\\
\hline
\hline
$\text{Re}(A)$& $\begin{aligned}
  \smash[b]{\vphantom{\bigg|}}
  -\cos2\theta_m +\bar{\gamma} \bar{\Delta}_b \sin2\theta_m \cos\chi
  \smash[t]{\vphantom{\bigg|}}
\end{aligned}$
\\
\hline
$\text{Im}(A)$&$\begin{aligned}
  \smash[b]{\vphantom{\bigg|}}
  -\bar{\gamma}\bar{\Delta}_a \sin2\theta_m \cos\chi
  \smash[t]{\vphantom{\bigg|}}
\end{aligned}$
\\
\hline
$|A|^2$&$\begin{aligned}
  \smash[b]{\vphantom{\bigg|}}
  \cos^2 2\theta_m - 2\bar{\gamma} \bar{\Delta}_b \sin 2\theta_m
  \cos2\theta_m \cos \chi
  \smash[t]{\vphantom{\bigg|}}
\end{aligned}$
\\
\hline
$|B|^2$&$\begin{aligned}
  \smash[b]{\vphantom{\bigg|}}
  \sin^2 2\theta_m +2\bar{\gamma}\sin 2\theta_m
  \Big(\bar{\Delta}_a \sin \chi+ \bar{\Delta}_b \cos 2\theta_m
  \cos\chi\Big)
  \smash[t]{\vphantom{\bigg|}}
\end{aligned}$
\\
\hline
\end{tabular}
\end{table}

It should be noted that in the two-flavor approximation
in the absence of neutrino decay,
i.e. $b_1=b_2=\gamma=0$,  we have $P_{\alpha \alpha} = P_{\beta\beta}$
and $P_{\beta \alpha} = P_{\alpha \beta}$. These equalities no longer hold in the
presence of decay.

\smallskip

\noindent {\it Neutrino flavor conversions
  up to $O(\bar\gamma^2)$} --- 
For probabilities accurate up to order $\bar\gamma^2$, we need to calculate the
term in eq.(\ref{eq:gammasq}) that involves a double summation.
This sum may be rewritten as
\begin{equation}
  \dfrac{1}{2}
  \sum\limits_{n_1=1}^\infty
  \left( \sum\limits_{n_2=1}^\infty
  \mathcal{Y}_{n_2}\mathcal{Y}_{n_1} +
  \sum\limits_{n_2=n_1}^\infty
  \frac{n_1-n_2}{n_1+n_2}
       [\mathcal{Y}_{n_2},\mathcal{Y}_{n_1}]\right),
\end{equation}
whose closed form may be obtained using the observation
\begin{equation}
  [\mathcal{Y}_{n_2},\mathcal{Y}_{n_1}]=\frac{(-1)^{n_2}-(-1)^{n_1}}{4n_1! n_2!}
  (i\Delta_d t)^{n_2+n_1-2} (\gamma t)^2\sigma_3 \;.
\end{equation}

The eigenvalues of ${\mathcal H}_m$ get corrections at $O(\bar\gamma^2)$,
and it is convenient to write the probabilities at this (and higher)
order in terms of the difference of the exact eigenvalues
\begin{equation}
    \Delta_D = \sqrt{\Delta_d^2-\gamma^2} \; .
\end{equation}
The probabilities at $O(\bar\gamma^2)$ can be written in the same
form as eqs.~(\ref{eq:PeeUptogamma}) and (\ref{eq:PemuUptogamma})
with the replacements
\beq
\Delta_a \to   {\rm Re}(\Delta_D) \; ,  \quad
\Delta_b  \to   - {\rm Im}(\Delta_D) \; ,
\label{eq:delta-Delta}
\eeq
and the entries in Table~\ref{tab:AmpTableGamma} replaced by
\barr
A(\chi) & \to &  A^{(0)} + \gamma A^{(1)} - \gamma^2
\cos 2\theta_m / (2 \Delta_d^2) \;,  \label{eq:a2} \\
B(\chi) & \to  &  A^{(0)} + \gamma A^{(1)} + \gamma^2
\sin 2\theta_m / (2 \Delta_d^2) \; , \label{eq:b2}
\earr
The entries corresponding to Table~\ref{tab:ProbTableGamma} can be calculated using
eqs.~(\ref{eq:a2}) and (\ref{eq:b2}).

\smallskip

\noindent {\it Exact results} --- 
For the 2-flavor system, it is also possible to obtain the exact expressions
for neutrino survival and conversion probabilities
by expressing $-i {\mathcal H}_m t$ as a linear combination of Pauli matrices
\cite{nielsen2002quantum}.
For any $2 \times 2$ matrix $\mathbb{K}$,
one can write
\begin{equation}
  e^{\mathbb{K}} = e^{k_0} \left[ \mathbb{I} ~ \cosh k
  + \frac{\vec{k} \cdot \vec{\sigma}}{k}  ~\sinh k \right] ,
\end{equation}
where $k_\mu \equiv {\rm Tr}(\mathbb{K} \cdot \sigma_\mu)/2$,
and $k \equiv \sqrt{k_1^2+k_2^2+k_3^2}$.
For the matrix $\mathbb{K} = -i {\mathcal H}_m t$ as in eq.~(\ref{eq:Hm-def}),
this corresponds to 
\beq
k_0 = - \frac{it}{2} (d_1 + d_2) \; , \;
k =\frac{i t \Delta_D}{2} \; .
\eeq
This leads to the exact probabilities that can be written in the same
form as eqs.~(\ref{eq:PeeUptogamma}) and (\ref{eq:PemuUptogamma}),
with the replacements given in eq.~(\ref{eq:delta-Delta}), and
\barr
A(\chi) \to  \frac{\Delta_d}{\Delta_D} A(\chi)  \quad  & , &  \quad 
B(\chi) \to   \frac{\Delta_d}{\Delta_D} B(\chi)  \;,
\label{eq:ab-exact}
\earr
in Table~\ref{tab:AmpTableGamma}.

The entries corresponding to Table~\ref{tab:ProbTableGamma} can be calculated using
eq.~(\ref{eq:ab-exact}).

\smallskip

\noindent {\it Numerical comparison} ---
We now demonstrate, using numerical calculations, 
the convergence of our analytical results towards the
exact neutrino oscillation probabilities, when higher and higher order
terms in $\bar\gamma$ are included.
For the sake of illustration, we choose the survival probability of
$\nu_\mu$ with energy  $E\sim$ GeV, for a baseline of 295 km.
This would correspond to the probability relevant at the T2K/T2HK
experiment.
\begin{figure*}[]
  \centering
  \includegraphics[width=0.48\textwidth]{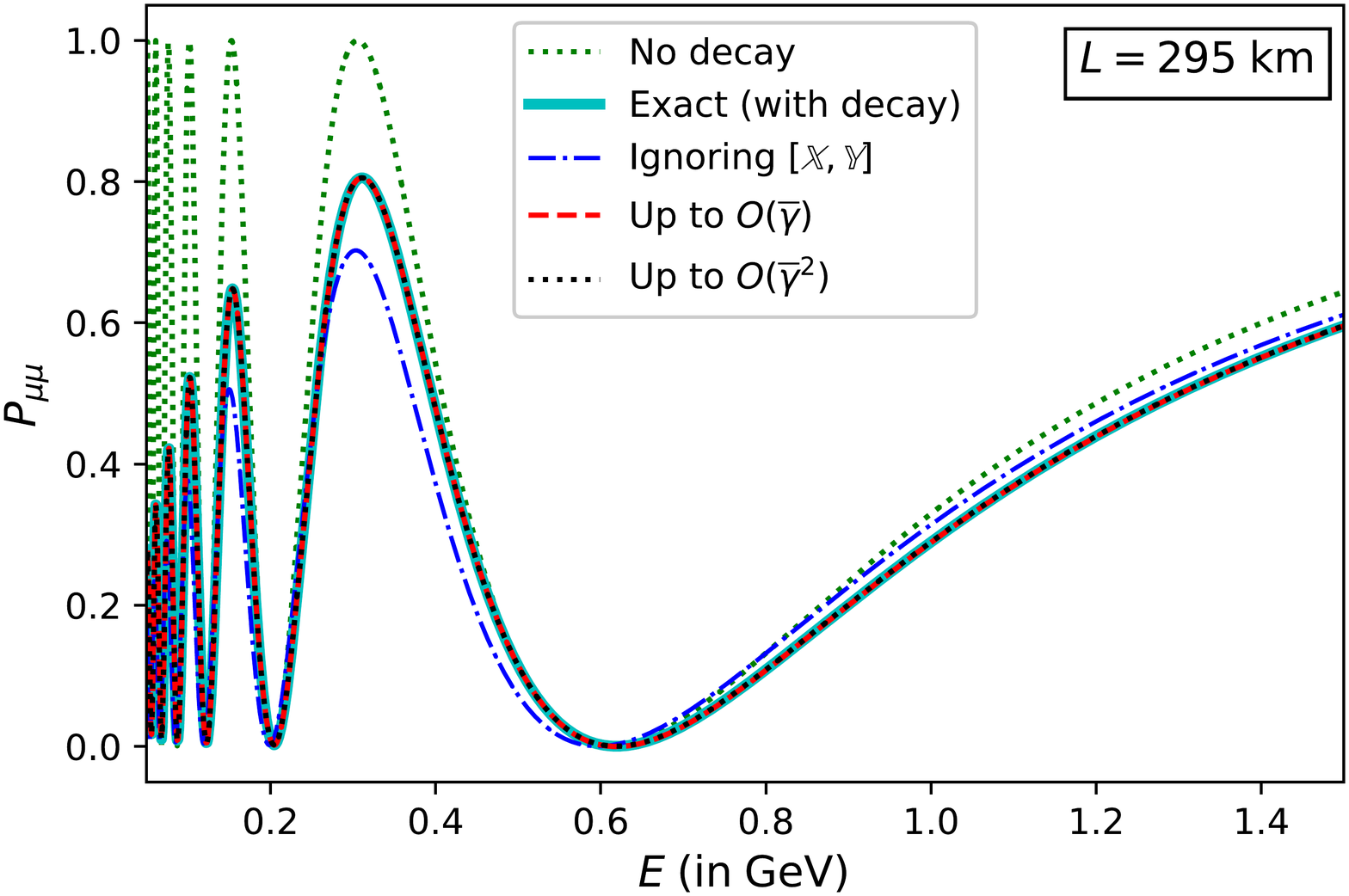}
  \includegraphics[width=0.49\textwidth]{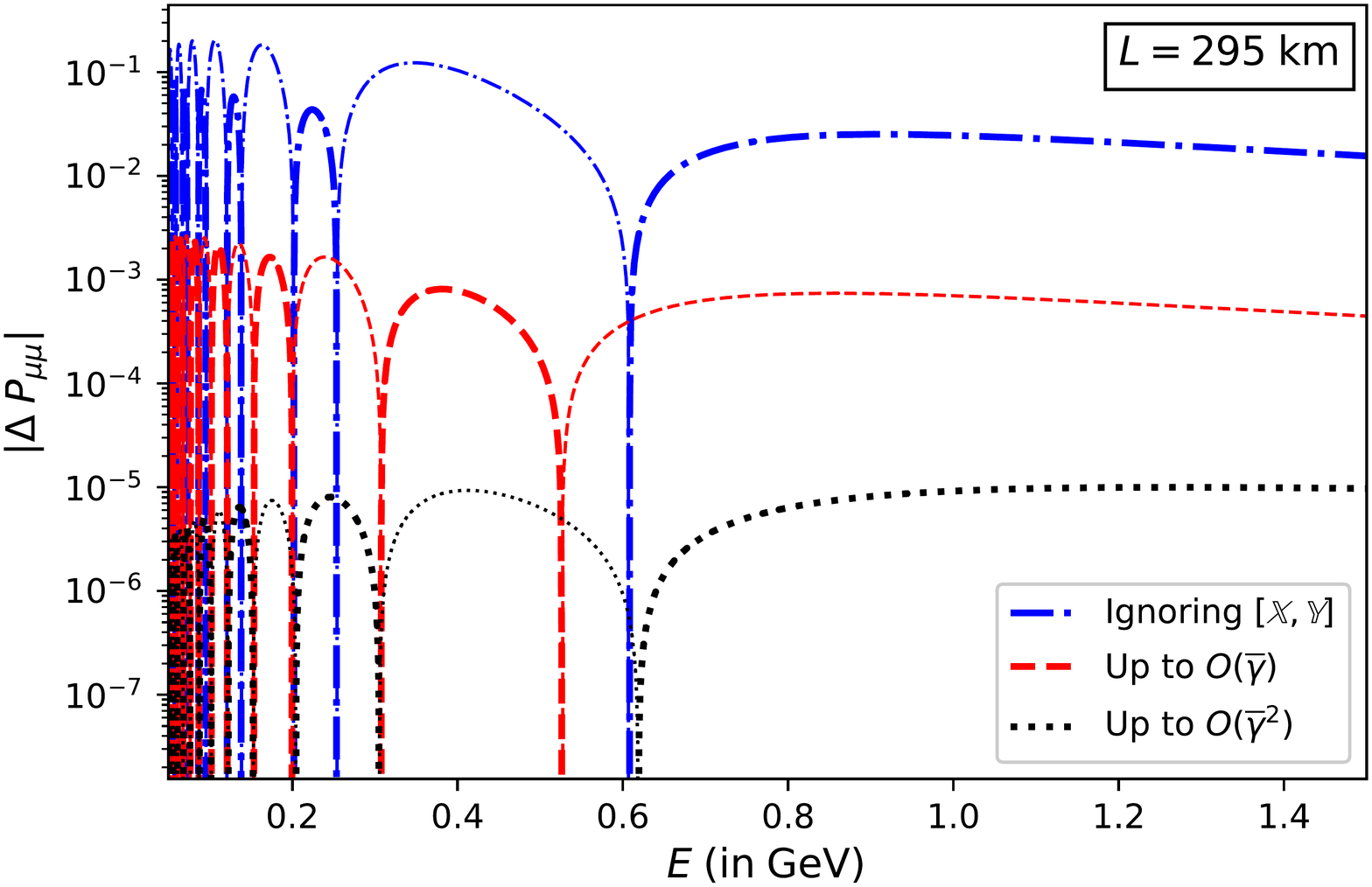}\\
  \caption{The left panel shows the survival probability $P_{\mu\mu}$
    calculated exactly, and by using the analytical expressions in the text,
    in the presence and absence of neutrino decay.
    The right panel shows the differences between the analytical expressions
    and the exact results for decay.
    The thick (thin) lines correspond to $\Delta P_{\mu\mu} >0$
    ($\Delta P_{\mu\mu} <0$). 
    The sharp dips in the right panel correspond to those energies where
    the analytical expressions give the same values as the exact ones.
  }
  \label{fig:probs}
\end{figure*}
We take the parameters of the Hermitian part of the Hamiltonian to be
 $ \Delta_a = 2.56 \times10^{-3} \text{ eV}^2 / (2E) ,
  \theta_m = 45 ^\circ$,
and the parameters of the anti-Hermitian part to be
$(b_1, b_2, \gamma) = (3,6,8) \times 10^{-5} \mbox{ eV}^2 / (2 E),\chi=\pi/4$.
Note that the parameters $a_i, b_i, \gamma$ from Eq.~(\ref{eq:Hm-def})
are taken to vary as $1/E$ to account for time dilation.

The left panel in Fig.~\ref{fig:probs} shows the probability $P_{\mu\mu}(E)$
without decay, and successive approximations at $O(\bar\gamma)$
and $O(\bar\gamma^2)$ in the presence of decay. The incorrect approximation
that neglects the commutator $[\mathbb{X},\mathbb{Y}]$ is also indicated.
The convergence towards the exact solution is more clearly demonstrated
in the right panel of Fig.~\ref{fig:probs}, where we show the values of
the error
\beq
|\Delta P_{\mu\mu}| \equiv P_{\mu\mu}\mbox{(analytical)} - P_{\mu\mu}\mbox{(exact)}
\; \eeq
on a logarithmic scale.
    Clearly, the inclusion of $O(\bar\gamma)$ and
    $O(\bar\gamma^2)$ terms reduces the error by orders of magnitude.

\noindent {\it Comparison with earlier results} ---
In ref.~\cite{Berryman:2014yoa}, neutrino decay in vacuum was analyzed
using diagonalization of the non-Hermitian Hamiltonian
with ${\mathcal H}_{\rm diag} = N^{-1} {\mathcal H} N$, 
using a non-unitary matrix $N$. 
We find that the most general form of the non-unitary matrix that would
diagonalize a non-Hermitian ${\mathcal H}$ is
\beq
N =  \left(
\begin{array}{cc}
 \cos \theta_m  & \sin \theta_m  \\
 -\sin \theta_m & \cos \theta_m \\
\end{array}
\right) \left(
\begin{array}{cc}
1 & -i \frac{\gamma  e^{i \chi }}{ \Delta_D + \Delta_d } \\
i \frac{ \gamma  e^{-i \chi }}{ \Delta_D + \Delta_d } & 1 \\
\end{array}
\right) \; .
\label{eq:N-mat}
\eeq
Note that since $(\Delta_D+\Delta_d)$ is complex,
the off-diagonal elements of the second matrix in eq.~(\ref{eq:N-mat})
are not complex conjugates of one another,
an assumption implicitly made in ref.~\cite{Berryman:2014yoa}.
This introduces corrections in the neutrino conversion probabilities
of $\sim O(\bar\gamma\bar\Delta_b)$. These may be neglected if one assumes
$\bar{\Delta}_b \sim O(\bar\gamma)$; however, these will then contribute to
$O(\bar\gamma^2)$ corrections.

For the special case with only one unstable neutrino, whose decay
eigenstate in vacuum coincides with one of the mass
eigenstates $\nu_2$ in vacuum, the probabilities in matter may be
obtained by the following identifications:
\barr
a_{1,2} = \frac{\tilde{m}_{1,2}^2}{2E} \quad & , & \quad  
b_{1,2} = \frac{\alpha_2}{4E}  [1 \mp \cos[2(\theta - \theta_m)] \; , \\
\chi=0 \quad & , & \quad
\gamma = \frac{\alpha_2}{2E} \sin[2(\theta-\theta_m)] .
\earr
Here, $\tilde{m}_i(m_i)$ and $\theta_m(\theta)$ are the mass eigenvalues
and mixing angle in matter (vacuum), and
$\alpha_2 = m_2 /\tau_2$, where $\tau_2$ is the
lifetime of $\nu_2$ in vacuum.
Note that both mass eigenstates in matter now undergo decay, and
the off-diagonal term $\gamma$ is generated,
even though it was absent in vacuum.
This prescription gives the correct analytical probability
expressions for decaying neutrinos in matter,
which are hitherto not explicitly given in the literature.
In the limit $\theta_m \to \theta$ and $\tilde{m}_i \to m_i$, the standard probabilities
for decay in vacuum \cite{Lindner:2001fx} are obtained.

\noindent {\it Concluding remarks}  --- 
Neutrino decay is characterized by a non-Hermitian Hamiltionian,
which cannot be diagonalized with a unitary transformation.
Further, there is no guarantee that decay eigenstates are
the same as the mass eigenstates, although it is usually assumed.
We point out that even if these two sets are the same in vacuum,
matter effects necessarily change this simple picture,
warranting a more careful treatment.

In this Letter, we  develop a novel formalism which can address  the
above two issues, and allows one to obtain compact
analytical forms for two flavour probabilities even in matter.
The crucial step in our formulation is to perform the analysis in the basis
of mass eigenstates in matter in the absence of decay, so that
the Hermitian component of the Hamiltonian is diagonal.
The anti-Hermitian decay matrix is not diagonal in this basis,
and does not commute with the Hermitian part.
This prompts us to employ the Zassenhaus (inverse BCH) expansion for the
time evolution matrix. 
Further, we introduce a resummation of commutators to compute the neutrino
survival and conversion probabilities perturbatively in $\bar\gamma$,
the parameter characterizing the mismatch between mass and decay eigenstates.
This is the first time such a formulation has been used to treat
propagation of unstable neutrinos in matter. 

While the explicit results in this work are calculated
in the context of a two-flavor
scenario, the framework of perturbative expansion in $\bar\gamma$
may be easily extended to three flavors.
Moreover, the scope of application of this method
goes beyond just the neutrino decay hypothesis; the formalism may be
applied to various other phenomena such as the combined treatment of oscillations and
absorption for high energy neutrinos, axion-photon
oscillations in an optically semi-opaque medium, or even the neutral
meson mixing systems. 

\smallskip
\noindent {\it Acknowledgments} ---
D.S.C. would like to thank S. Moitra for useful discussions.
S.G. and L.S.M. would like to thank S. Choubey and C. Gupta for useful discussions.
A.D. acknowledges support from the Department of Atomic Energy (DAE),
Government of India, under Project Identification No. RTI4002. 
S.G.  acknowledges  the J.C Bose Fellowship  (JCB/2020/000011)
of Science and  Engineering Research Board  of Department of Science
and Technology, Government of India. 

\end{document}